\documentclass{ws-mpla2}
\usepackage{float}
\usepackage{multirow}
\usepackage{xspace}
\usepackage{relsize}

\newcommand{\gev}{\ensuremath{\mathrm{\,Ge\kern -0.1em V}}\xspace}
\newcommand{\mev}{\ensuremath{\mathrm{\,Me\kern -0.1em V}}\xspace}
\newcommand{\kev}{\ensuremath{\mathrm{\,ke\kern -0.1em V}}\xspace}

\def\babar{\mbox{\slshape B\kern-0.1em{\smaller A}\kern-0.1em B\kern-0.1em{\smaller A\kern-0.2em R}}}

\def\epem{e^+e^-}

\def\lplm{l^+l^-}

\def\CP{\ensuremath{C\!P}\xspace}

\begin{document}

\markboth{B. Echenard}{Search for Low-Mass Dark Matter at \babar}

%%%%%%%%%%%%%%%%%%%%% Publisher's Area please ignore %%%%%%%%%%%%%%
\catchline{}{}{}{}{}
%%%%%%%%%%%%%%%%%%%%%%%%%%%%%%%%%%%%%%%%%%%%%%%%%%%

\title{Search for Low-Mass Dark Matter at \babar}

\author{\footnotesize Bertrand Echenard}

\address{Department of Physics, California Institute of Technology, MC 356-48\\
Pasadena, California 91125,USA\\
echenard@caltech.edu}

\maketitle

%\pub{Published in Mod. Phys. Lett. A, Vol. 27, No. 18 (2012) 1230016, 
%doi: 10.1142/S0217732312300169 \copyright copyright World Scientific Publishing Company http://www.worldscientific.com}{}

\begin{abstract}

This review briefly describes light dark matter searches performed 
by the \babar\ experiment. Although dark matter candidates have traditionally 
been associated with heavy particles appearing in extensions of the Standard 
Model, a lighter component remains a well motivated alternative, and many 
scenarios of light dark matter have been recently proposed. Thanks to their large 
luminosities, $B$ factories offer an ideal environment to probe these 
possibilities, complementing searches from direct detection and satellite 
experiments.

\keywords{low-mass dark matter; Upsilon decays, hidden sector; $B$ factories.}
\end{abstract}

\ccode{PACS Nos.: 12.60-i ,13.25.Gw,14.80.Ec,14.80.Da}

\section{Introduction}	

The evidence for dark matter is now overwhelming, and 
the determination of its properties is one of the central tasks in 
current theoretical and experimental investigations. Of the vast 
array of dark matter candidates proposed during the last decades, 
the class known as Weakly Interacting Massive Particles 
(WIMPs) has many attractive features. 
Although WIMPs are traditionally associated with heavy particles 
appearing in models of New Physics, a lighter component remains 
a theoretically well-motivated alternative.

Many models have been recently proposed to accommodate light 
dark matter, ranging from minimal setups including a single 
dark matter particle to hidden sectors with arbitrary 
structures. Thanks to their large luminosities and 
low-background environments, $B$ factories offer an ideal 
environment to probe for $\mev$-$\gev$ dark matter, complementing 
searches from the LHC, direct detection and satellite 
experiments. 

During the last decade, the \babar\ experiment~\cite{Bib:Babar} has 
collected more than $500 \rm \, fb^{-1}$ of data at the 
$\Upsilon(4S), \Upsilon(3S),\Upsilon(2S)$ resonances and just below 
the $\Upsilon(4S)$ resonance. This large dataset has been exploited to 
shed some light on many aspects of precision physics, from \CP 
violation to spectroscopy, and searches for light New Physics. In 
the following, we present a review of searches for low-mass dark 
matter at \babar.

\section{Search for light dark matter in invisible $\Upsilon(1S)$ decays.}

In a minimal scenario, a single dark matter field ($\chi$) is added to the 
Standard Model (SM), together with a new boson that mediates SM-dark matter  
interactions~\cite{McElrath:2005bp,Yeghiyan:2009xc,Fayet:2009tv}. This 
boson could be produced in $b\bar{b}$ annihilation, and subsequently 
decay to a pair of dark matter particles, contributing to the invisible 
width of the $\Upsilon$ mesons. In the SM, invisible $\Upsilon(1S)$ decays proceed 
through the $\nu \bar{\nu}$ final state with a branching fraction 
$B(\Upsilon(1S) \rightarrow \nu \bar{\nu}) \sim (1 \times 10^{-5})$ 
\cite{Chang:1997tq}, well below the current experimental sensitivity. Assuming no flavor-changing 
currents, calculations based on the thermal dark matter relic density 
predict a rate $\Upsilon(1S) \rightarrow \chi \bar{\chi}$ larger by 
one or two orders of magnitude than that of 
$\Upsilon(1S) \rightarrow \nu \bar{\nu}$~\cite{McElrath:2007sa}.

A search for dark matter in invisible $\Upsilon(1S)$ decays has been 
performed at $\babar$ using a sample of $122\times 10^6$ $\Upsilon(3S)$ 
mesons~\cite{Aubert:2009ae}. A clean sample of 
$\Upsilon(1S)$ mesons is obtained by reconstructing   
$\Upsilon(3S) \rightarrow \pi^+\pi^- \Upsilon(1S)$ transitions. 
The signal topology consists of exactly two oppositely-charged tracks 
originating from the interaction point with no additional activity in 
the detector. The events are selected using a multivariate classifier 
based on variables describing the pions, the neutral energy deposited 
in the calorimeters and the multiplicity of $K^0_L$ candidates. 

\begin{figure}[htb]
\begin{center}
\includegraphics[width=0.6\textheight]{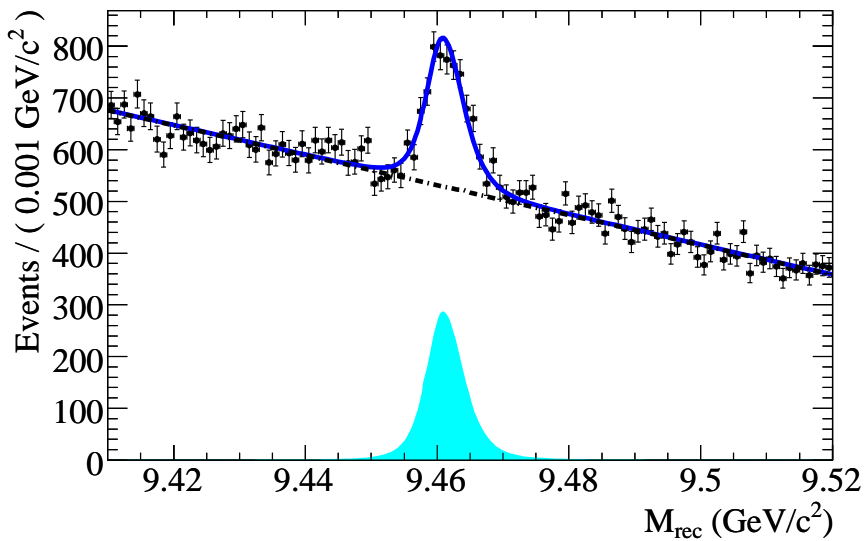} 
\end{center}
\caption{The distribution of the dipion recoil mass ($M_{rec}$) together with the 
result of the maximum likelihood fit (full line). The non-resonant background (dashed line) 
and the sum of the signal and the peaking background (solid filled) are also shown.}
\label{Fig::inv1}
\end{figure}

The distribution of the resulting dipion recoil mass, shown in Fig.~\ref{Fig::inv1}, 
exhibits a clear peak corresponding to $\Upsilon(1S)$ mesons, on top 
of a non-resonant component. In addition to signal events, a peaking 
background arising mainly from $\Upsilon(1S)$  two-body decays, where 
the decay products escape undetected, is also present. This component, 
which is kinematically indistinguishable from signal events, is estimated 
from Monte Carlo simulations. After subtraction of the peaking 
background, the signal yield is found to be $-118 \pm 105 \pm 124$, 
where the first uncertainty is statistical and the second systematic. 

Correcting for efficiency, a branching fraction 
$B(\Upsilon(1S) \rightarrow \rm invisible) = (-1.6 \pm 1.4 \pm 1.6) \times 10^{-4}$ is obtained. Lacking 
evidence for such decays, a 90\% confidence level Bayesian upper limit on its 
branching fraction is set at $3.0 \times 10^{-4}$  using priors flat in branching 
fraction. This result improves the best previous measurement~\cite{Tajima:2006nc} 
by almost an order of magnitude, and sets stringent constraints on minimal light 
dark matter models.

\section{Search for light dark matter in radiative $\Upsilon(1S)$ decays.}

In addition to purely invisible decays, light dark matter can  
be probed in radiative $\Upsilon \rightarrow \gamma \chi \bar{\chi}$ decays. The 
corresponding rate is suppressed by a factor ${\cal O}(\alpha)$ compared to 
that of the pure invisible decay, and a branching fraction in the range 
$10^{-5} - 10^{-4}$ is expected~\cite{Yeghiyan:2009xc}.

This final state is also sensitive to dark matter candidates predicted by several 
supersymmetric extensions of the Standard Model.
Although the Minimal Supersymmetric Standard
Model (MSSM) cannot support light dark matter, because it would require another charged 
or colored particle to be light as well, trivial extensions of the 
MSSM can easily evade these constraints. For example, the Next-to-Minimal 
Supersymmetric Standard Model includes the possibility of a neutralino ($\chi_0$)  and a 
\CP-odd Higgs boson ($A^0$) at the $\gev$ scale~\cite{Gunion:2005rw}. If kinematically 
allowed, the decay $\Upsilon \rightarrow \gamma + \rm \, invisible$ could proceed 
through Wilczek production of a light scalar~\cite{Wilczek:1977zn}, followed by the 
decay into a pair of neutralinos: $\Upsilon \rightarrow \gamma A^0, A^0 \rightarrow 
\chi_0 \bar{\chi}_0$. The branching fraction $\Upsilon \rightarrow \gamma A^0, 
A^0 \rightarrow \chi_0 \bar{\chi}_0$ 
is predicted~\cite{Gunion:2005rw} to be as large as $10^{-4} - 10^{-3}$, making 
it accessible to $B$ factories.

A search for this process in radiative $\Upsilon(1S)$ decays has been performed 
at $\babar$ based on a sample of $99\times 10^6$ $\Upsilon(2S)$ mesons~\cite{delAmoSanchez:2010ac}. 
The $\Upsilon(1S)$ mesons are tagged using the dipion transition $\Upsilon(2S) 
\rightarrow \pi^+\pi^- \Upsilon(1S)$. The signal signature is a 
photon with two oppositely charged tracks identified as pions. 
The signal is extracted by a series of bidimensional unbinned likelihood fits 
to the dipion recoil mass and the missing mass squared. Values of $m_{\chi_0}$ 
are probed in steps of $0.1 - 0.5 \gev$ over $0 \leq m_{\chi_0} \leq 4.5 \gev$. 

\begin{figure}[htb]
\begin{center}
\includegraphics[width=0.6\textheight]{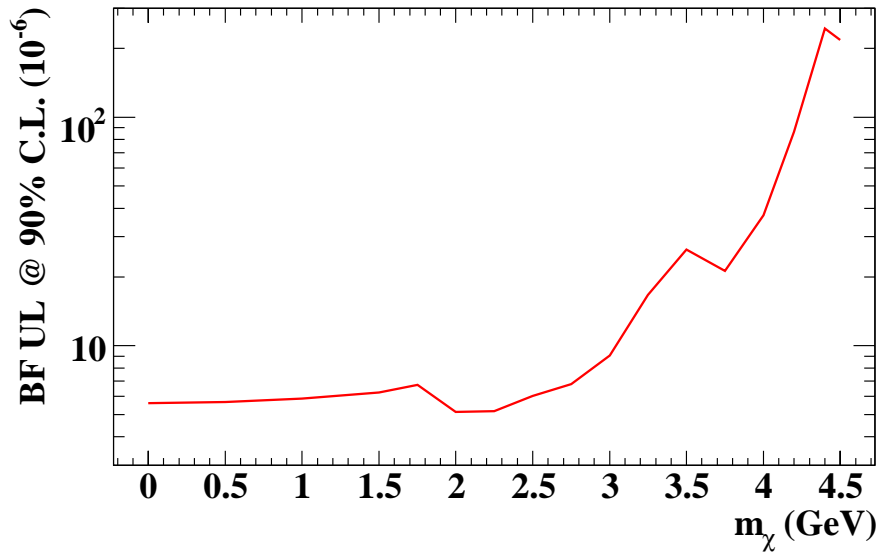} 
\end{center}
\caption{90\% confidence level upper limit on the branching fraction 
$B(\Upsilon(1S) \rightarrow \gamma \chi_0 \bar{\chi}_0)$ as a function of $m_{\chi_0}$.}
\label{Fig::rad1}
\end{figure}

No significant signal is observed, and 90\% confidence level limits on the branching 
fraction $B(\Upsilon(1S) \rightarrow \gamma \chi_0 \bar{\chi}_0)$ are set in the range
 $(0.5-24) \times 10^{-5}$ (Fig.~\ref{Fig::rad1}), assuming a phase-space distribution for the 
photon energy. This result improves the best previous measurement~\cite{Balest:1994ch} by almost 
two orders of magnitude, and constrains on light Higgs boson and dark matter models.

\section{Search for dark matter and hidden sectors}

A new class of dark matter model has recently been proposed, motivated 
by results from terrestrial and satellite experiments. These theories 
introduce a new hidden sector with WIMP-like fermionic dark matter particles 
charged under a new Abelian gauge group~\cite{Fayet,Pospelov:2007mp,ArkaniHamed}. The corresponding gauge 
boson, dubbed a hidden photon ($A'$), couples to SM fermions through its 
kinetic mixing with the SM photon~\cite{Holdom}.
The hidden photon mass is constrained to be at most a few $\gev$ to be 
compatible with the electron/positron excess observed by 
PAMELA~\cite{Adriani:2008zr} and FERMI~\cite{Abdo:2009zk,Ackermann:2010ij}, 
without a comparable anti-proton signal. 

The hidden boson masses are usually generated via the Higgs 
mechanism, adding hidden Higgs bosons $(h')$ to the theory. A minimal hidden sector 
model includes a single hidden photon and a Higgs boson~\cite{Batell:2009yf}. 
Additional Higgs and gauge bosons appear in more complex setups~\cite{Essig:2009nc}.
The coupling strength between the SM and the hidden sector, $\epsilon$, can 
naturally be as large as  $10^{-4} - 10^{-2}$~\cite{Batell:2009yf}, opening the 
possibility of probing a light hidden sector at low-energy $\epem$ colliders.

\subsubsection{Search for a hidden photon}

\begin{figure}
\begin{center}
\includegraphics[width=0.65\textwidth]{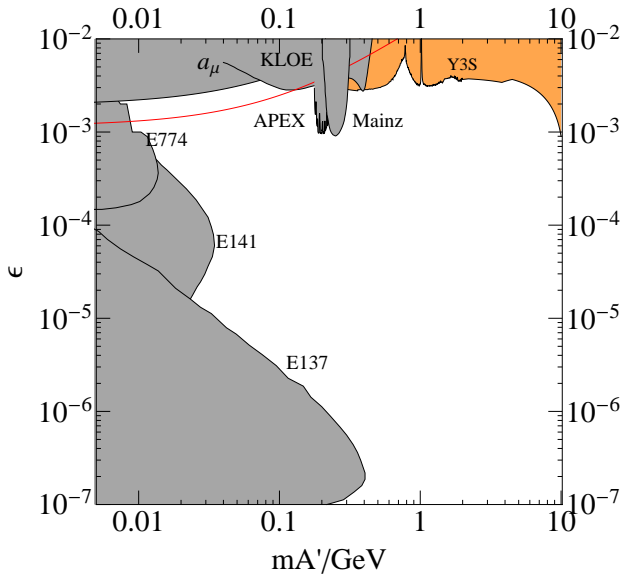}
\caption{Constraints on the mixing parameters, $\epsilon$, as a function of the 
hidden photon mass derived from searches in $\Upsilon(2S,3S)$ decays at \babar\ 
(orange shading) and from other experiments 
\protect\cite{Bjorken:2009mm,Giovannella:2011nh,Abrahamyan:2011gv} (gray shading). 
The red line shows the value of the coupling required to 
explain the discrepancy between the calculated and measured anomalous magnetic 
moment of the muon~\protect\cite{Pospelov:2008zw}.}
\label{Fig::hp}
\end{center}
\end{figure}

Hidden photons can be readily produced in $\epem \rightarrow \gamma A', A' \rightarrow l^+l^-$ 
($l=e,\mu$) interactions, and be detected as resonances in the 
$l^+l^-$ spectrum. This signature is similar to that of light \CP-odd Higgs 
production in $\epem \rightarrow \Upsilon(2S,3S) 
\rightarrow \gamma \mu^+\mu^-$~\cite{Aubert:2009cp}, and searches for 
this final state have been reinterpreted as searches for hidden photon 
production~\cite{Bjorken:2009mm}. These limits are shown in Fig.~\ref{Fig::hp}, 
together with the constraints derived from the measurement of the muon anomalous 
magnetic moment, and searches at low-energy $\epem$ colliders and 
fixed-target experiments~\cite{Bjorken:2009mm,Giovannella:2011nh,Abrahamyan:2011gv}. 
Values of the mixing strength down to $10^{-3} - 10^{-2}$ are excluded for 
$0.212 < m_{A'} < 9.3 \gev$. Future analyses of the $\epem \rightarrow \gamma \lplm$ ($l=e,\mu$) 
final states are expected to probe values of the mixing strength down to $10^{-3}$ 
in the range $0.01 < m_{A'} < 10 \gev$, significantly improving existing bounds.

Other measurements may be reinterpreted as constraints on hadronic hidden 
photon decays. These include searches for peaks in inclusive $\epem \rightarrow \gamma +\rm{hadrons}$ 
production~\cite{Lees:2011wb}, as well as exclusive hadronic final states, 
such as $\epem \rightarrow \gamma \pi^+\pi^-$~\cite{Aubert:2009ad}. Invisible 
decays, which occur if all hidden sector 
states decay to long-lived particles, could also be detected as a mono-energetic 
photon signature in $\epem \rightarrow \gamma +\rm{invisible}$ 
events~\cite{arXiv:0808.0017}.

\subsubsection{Search for hidden gauge bosons}

Extensions of hidden sectors to a non-Abelian groups introduce additional 
hidden gauge bosons to the theory, generically denoted $W',W'',...$. The detailed 
phenomenology depends on the structure of the model, but heavy hidden 
bosons decay to lighter hidden states if kinematically accessible, while the 
lightest bosons are metastable and decay to SM fermions via their mixing 
with the hidden photon~\cite{Essig:2009nc}.

A pair of light hidden bosons can be produced via an off-shell $A'$ in 
$\epem \rightarrow A'^* \rightarrow W' W''$ collisions. 
\babar\ performed a search for di-boson production using 513 
fb$^{-1}$ of data, collected 
mostly at the $\Upsilon(4S)$ resonance~\cite{arXiv:0908.2821}. 
The bosons are reconstructed via their decay into muons or 
electrons. It is furthermore assumed that their mass is nearly 
degenerate\footnote{The production of identical bosons is forbidden by the 
Landau-Yang theorem}, and their width is negligible compared 
to the detector resolution.

The signature consists of four leptons with zero total charge carrying 
the full beam energy, forming two narrow dileptonic resonances with 
similar masses and width consistent with the experimental resolution. 
This topology is quite unique; the only backgrounds arise from 
QED processes. 

The signal is extracted as a function of the average dileptonic mass 
in the range $0.24 - 5.3 \gev$ in $10 \mev$ steps. No significant signal 
is observed, and limits on the $\epem \rightarrow A'^* \rightarrow W' W''$ 
cross-section are set, assuming the two bosons have nearly degenerate 
masses and equal branching fractions into electrons and muons. 
The results are translated into limits on the product 
$\alpha_D \epsilon^2$, where $\alpha_D = g_D^2/4\pi$ and $g_D$ is 
the hidden sector gauge coupling constant. Values down to $10^{-10}$ 
can be probed. Assuming $\alpha_D = \alpha \simeq 1/137$, limits on the mixing strength 
at the level of $10^{-4} - 10^{-2}$ are set for this process.

\subsubsection{Search for hidden Higgs bosons}

The Higgsstrahlung process, $\epem \rightarrow A' h', h' \rightarrow A' A'$, 
offers another promising gateway to hidden sectors, since it is one of the few 
processes suppressed by a single power  of the mixing strength, and the 
background is expected to be small. The event topology depends on the boson 
mass: Higgs bosons heavier than two hidden photons decay promptly, while their lifetime 
becomes large enough to escape undetected in the regime $m_{h'} < m_{A'}$.

$\babar$ performed a search for a hidden Higgs boson in Higgsstrahlung production in the prompt 
decay regime based on 521 fb$^{-1}$ of data~\cite{Lees:2012ra}. The measurement is 
performed in the range $0.8 < m_{h'} < 10.0 \gev$ and $0.25 < m_{A'} < 3.0 \gev$, with 
the constraint $m_{h'} > 2 m_{A'}$. 
The signal is either fully reconstructed into lepton or pion pairs (exclusive mode), or partially 
reconstructed (inclusive mode). Only two of the three hidden photons are reconstructed in the latter 
case, and the four-momentum of the remaining hidden photon is identified to that of the recoiling 
system. The topology of the exclusive modes consists of six tracks having an invariant mass close to the 
center-of-mass energy, forming three hidden photon candidates with equal masses. Inclusive modes are 
first identified by selecting two dileptonic resonances with similar mass, and requiring the mass of 
the recoiling system to be compatible with the Higgsstrahlung hypothesis. 

No significant signal is observed. Using uniform priors in the cross-section, upper limits on the 
$e^+ e^- \rightarrow  A' h', h' \rightarrow A' A'$ cross section are derived as a 
function of the hidden Higgs and hidden photon masses, including systematic uncertainties.

\begin{figure}[!htb]
\begin{center}
\includegraphics[width=0.8\textwidth]{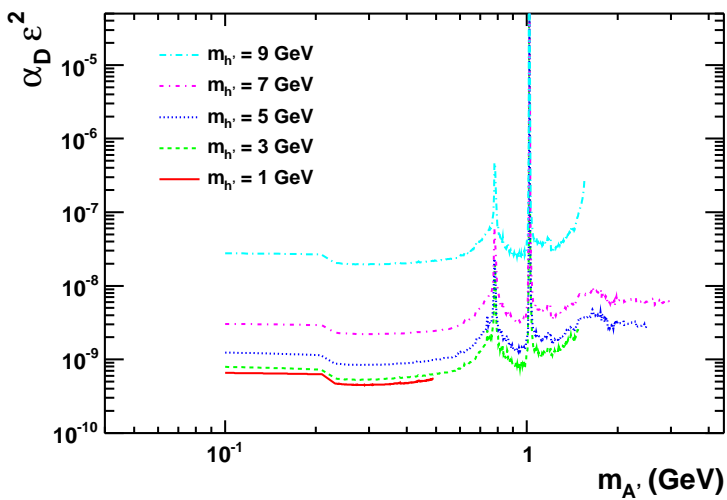} 
\caption{The 90\% confidence level upper limit on the product $\alpha_D \epsilon^2$ as a function 
of the hidden photon mass ($m_{A'}$) for selected values of hidden Higgs boson masses ($m_{h'}$).}
\label{Fig::higgs2}
\end{center}
\end{figure}

These results are translated into limits on the product $\alpha_D \epsilon^2$, and displayed in 
Fig.~\ref{Fig::higgs2} as a function of the hidden photon mass for selected values 
of the hidden Higgs boson mass. Values down to $10^{-10} - 10^{-8}$ are excluded for a 
large range of hidden photon and hidden Higgs masses, assuming prompt decays. Assuming 
$\alpha_D = \alpha \simeq 1/137$, limits on the mixing strength in the range 
$10^{-4} - 10^{-3}$ are derived, an order of magnitude smaller than the current experimental 
bounds extracted from direct photon production in this mass range.

\subsection{Conclusion}

Recent evidence has suggested that dark matter might contain a $\mev$-$\gev$ scale 
component. Thanks to their large luminosities, $B$ factories provide an ideal 
environment to probe for such a possibility, complementing direct detection and satellite 
experiments. No sign of light dark matter has been observed so far, and 
constraints on theoretical models have been set, significantly improving bounds 
from previous experiments. Super flavor factories are expected to increase the 
sensitivity of these searches by a factor $10-100$, further constraining the 
parameter space of these theories. Many more results are to come in the 
near future, which will hopefully shed some light on the nature of dark matter.

\section*{Acknowledgments}

I would like to thank David Hitlin and Yury Kolomensky for their comments on 
this manuscript, and Rouven Essig for useful theoretical discussions. I am also 
grateful to Matthew Graham for discussing the constraints on dark photon production 
and providing the corresponding figure. BE is supported by the Department of Energy, under 
grant DE-FG02-92ER40701.

\end{document}